# Adaptive Subcarrier and Bit Allocation for Downlink OFDMA System with Proportional Fairness

Mr. Sudhir B. Lande[1], Dr. J. B. Helonde[2], Dr. Rajesh Pande[3] S.S.Pathak[4]

[1] Asst. Prof. K.I.T.S. Ramtek Dist- Nagpur , India
[1] landeeed@yahoo.co.in

[2] Principal I T M Kamthee Dist- Nagpur , India
,[2] helondejb@yahoo.com

[3] Prof. & Vice-Principal R. C. O. E. M. Nagpur, India
[3] panderaj@yahoo.com

[4] Prof. in E&ECE dept. IIT Kharagpur. India
[4] sspathak@iit.ernet.in

## ABSTRACT

*This paper investigates the adaptive subcarrier and bit allocation algorithm for OFDMA systems. To minimize overall transmitted power, we propose a novel adaptive subcarrier and bit allocation algorithm based on channel state information (CSI) and quality state information (QSI). A suboptimal approach that separately performs subcarrier allocation and bit loading is proposed. It is shown that a near optimal solution is obtained by the proposed algorithm which has low complexity compared to that of other conventional algorithm. We will study the problem of finding an optimal sub-carrier and power allocation strategy for downlink communication to multiple users in an OFDMA based wireless system. Assuming knowledge of the instantaneous channel gains for all users, we propose a multiuser OFDMA subcarrier, and bit allocation algorithm to minimize the total transmit power. This is done by assigning each user a set of subcarriers and by determining the number of bits and the transmit power level for each subcarrier. The objective is to minimize the total transmitted power over the entire network to satisfy the application layer and physical layer. We formulate this problem as a constrained optimization problem and present centralized algorithms. The simulation results will show that our approach results in an efficient assignment of subcarriers and transmitter power levels in terms of the energy required for transmitting each bit of information, to address this need, we also present a bit loading algorithm for allocating subcarriers and bits in order to satisfy the rate requirements of the links*.

## KEYWORDS

*Orthogonal frequency division multiplexing, OFDMA, QSI, QoS, Dynamic resource allocation, Water-filling, SNR.*

## 1. INTRODUCTION

Orthogonal Frequency Division Multiple Access (OFDMA) is an attractive multiple access scheme for future wireless and mobile communication systems, which has been developed to support a variety of multimedia applications with different Quality-of-Service (QoS) requirements. OFDMA builds on Orthogonal Frequency Division Multiplexing (OFDM), which is immune to inter-symbol interference and frequency selective fading, as it divides the frequency band into a group of mutually orthogonal subcarriers, each having a much lower bandwidth than the coherence bandwidth of the channel. OFDMA can be readily implemented in OFDM systems by assigning each user a set of subcarriers. This allocation can be performed in a number of ways. The simplest method is a static allocation of subcarriers to each user. An improvement upon static allocation is dynamic subcarrier allocation, based on channel state conditions OFDMA is essentially a hybrid of FDMA and TDMA: Users are dynamically assigned subcarriers (FDMA) in different time slots (TDMA).The advantages of OFDMA start with the advantages of single-user OFDM in terms of robust multipath suppression and frequency diversity. In addition, OFDMA is a flexible multiple-access technique that can accommodate many users with widely varying applications, data rates, and QoS requirements. Because the multiple access is performed in the digital domain, before the IFFT operation, dynamic and efficient bandwidth allocation is possible[1,2,3]. This allows sophisticated time- and frequency- domain scheduling algorithms to be integrated in order to best serve the user population

## 1.1 OFDMA Overview

OFDMA technology takes an available wideband channel and divides it in the frequency domain into multiple narrower channels, each with its own subcarrier . This frequency division is assumed to be fine enough such that these subchannels experience only flat (narrowband) fading. Users in an OFDMA system are each assigned distinct and non-overlapping subsets of the subcarriers, and communicate with the base station over their dedicated set of subchannels.
The complexity arises in the allocation of subcarriers to users. Firstly, users in an OFDMA
system will generally experience different fading environments, and thus the signal-to-noise ratio (SNR) and received power variations seen by one user will differ from that seen by another. Secondly, different users have different QoS requirements, dictating the SNRs and received powers required at their receivers. Thirdly, there may either be total power constraints or requirements to minimize consumed power. Lastly, wireless channels are subject to constant change in real life, so resource allocation techniques must be fast enough to keep up with changing user environments if real-time applications are to be feasible. Now, in any multi-user system, there is a distinction between the uplink and the downlink. The former describes user-to-base-station communications, whereas the latter describes base-station-to-user communications[4]. These two cases are different due to fundamental differences between these channel models. As a result, resource allocation algorithms particularly those for power will be different for these two cases. Power allocation algorithms will not be presented unless inseparable from the subcarrier allocation algorithms, but even then, the paper will focus only on analysis of the subcarrier allocations.

## 1.2 Multiuser Diversity and Adaptive Modulation

In OFDMA, the subcarrier and the power allocation should be based on the channel conditions in order to maximize the throughput. In this section, we provide necessary background discussion on the key two principles that enable high performance in OFDMA: multiuser diversity and adaptive modulation. Multiuser diversity describes the gains available by selecting a user or sub- set of users having "good" conditions. Adaptive modulation is the means by which good channels can be exploited to achieve higher data rates. In multi-user environment, each user is dynamically assigned to a subset of subcarriers in each frame, to take advantage of the fact that at any time instance, channel responses are different for different users and at different subcarriers. This capability of OFDMA systems enables the network to perform a flexible radio resource management, such as dynamic subcarrier assignment (DSA), adaptive power allocation (APA), and adaptive modulation and coding (AMC) scheme to improve the system performance significantly under different traffic loads and time-varying channel conditions. We consider multiuser systems where multiple users are allowed to transmit simultaneously on different subcarriers per OFDM symbol. Mobile users on certain OFDM sub-channels may experience deep frequency-selective fading in a multipath propagation environment. Since each user may have a different sub-channel impulse response, a poor sub-channel for one user may be a good sub-channel for another user. Clearly, if a user who suffers from poor sub-channel gain can be reassigned to a better sub channel the total throughput can be increased. This is also known as multi-user diversity. Spatial reuse of channels, performed to increase the network capacity and conserve scarce network resources, should consider physical layer characteristics. Also, the medium access control (MAC) scheme needs to be sensitive to the channel quality experienced by different transmitters and multiuser diversity is exploited [5,6]. The goal of our work is to study the joint optimization of the physical and MAC layers. However, we will focus on subcarrier allocation and bit loading for OFDMA. Orthogonal Frequency Division Multiplexing (OFDM) is a spectrally efficient digital modulation technique. The spectrum of interest is divided into a number of parallel orthogonal narrowband sub-channels (known as subcarriers). When different subcarriers experience different channel gains and interference levels, adaptive bit loading is performed to enhance the throughput for a given transmission power limit.

## 2. PROBLEM FORMULATION

The aim is to allocate subcarrier at the bit loading algorithm using OFDMA for real time traffic. Here our objective is to assign a set of subcarriers to each link and load bits on each assigned subcarrier. The factors to be considered are the frequency-dependent path loss parameters and the interference due to other links that have already been assigned those subcarriers. Our algorithms are general and will be design to work for satisfying application and physical layer by improving CSI and QSI. Subcarrier allocation is performed at the base station and the users are notified of the carriers chosen for them. After the allocation, each user performs power allocation and bit loading across the subcarriers allocated to it to find the transmission power In this paper we will examine algorithms which fall between these three extremes. Intuitively, we are separating the problem into three stages:

- **Subcarrier Allocation:** we are going through an adaptive multiuser subcarrier allocation scheme where the subcarriers are assigned to the users based on instantaneous channel information.
- **Power Allocation:** Using the channel information, the transmitter applies the combined subcarrier, bit, and power allocation algorithm to assign different subcarriers to different users and the number of bits/OFDM symbol to be transmitted on each subcarrier.
- **Bit allocation:** By this optimal allocation of bits the overall transmit power will be minimized.

This paper is organized as follows: The section 3 deals system model of OFDMA systems. Certain existing subcarrier and power allocation algorithm are analyzed and presented in section 4 Section 5 deals with bit allocation algorithms for OFDM system. Section 6 deals with proposed subcarrier & sub-power allocation algorithm for OFDMA systems and section 7 analyses results of the proposed algorithm

## 3. OFDMA SYSTEM MODEL

A schematic diagram of an OFDMA-based system used in this paper is shown in Fig 1. At the transmitter, the serial data streams from the users are fed into the OFDM transmitter block. Using the channel information from all users, the subcarrier and bit/power allocation algorithm is applied to assign different subcarriers to different users. Here, it is assumed that a subcarrier at a particular time is not being shared among users. The power allocated to each subcarrier is also determined in the process[6].

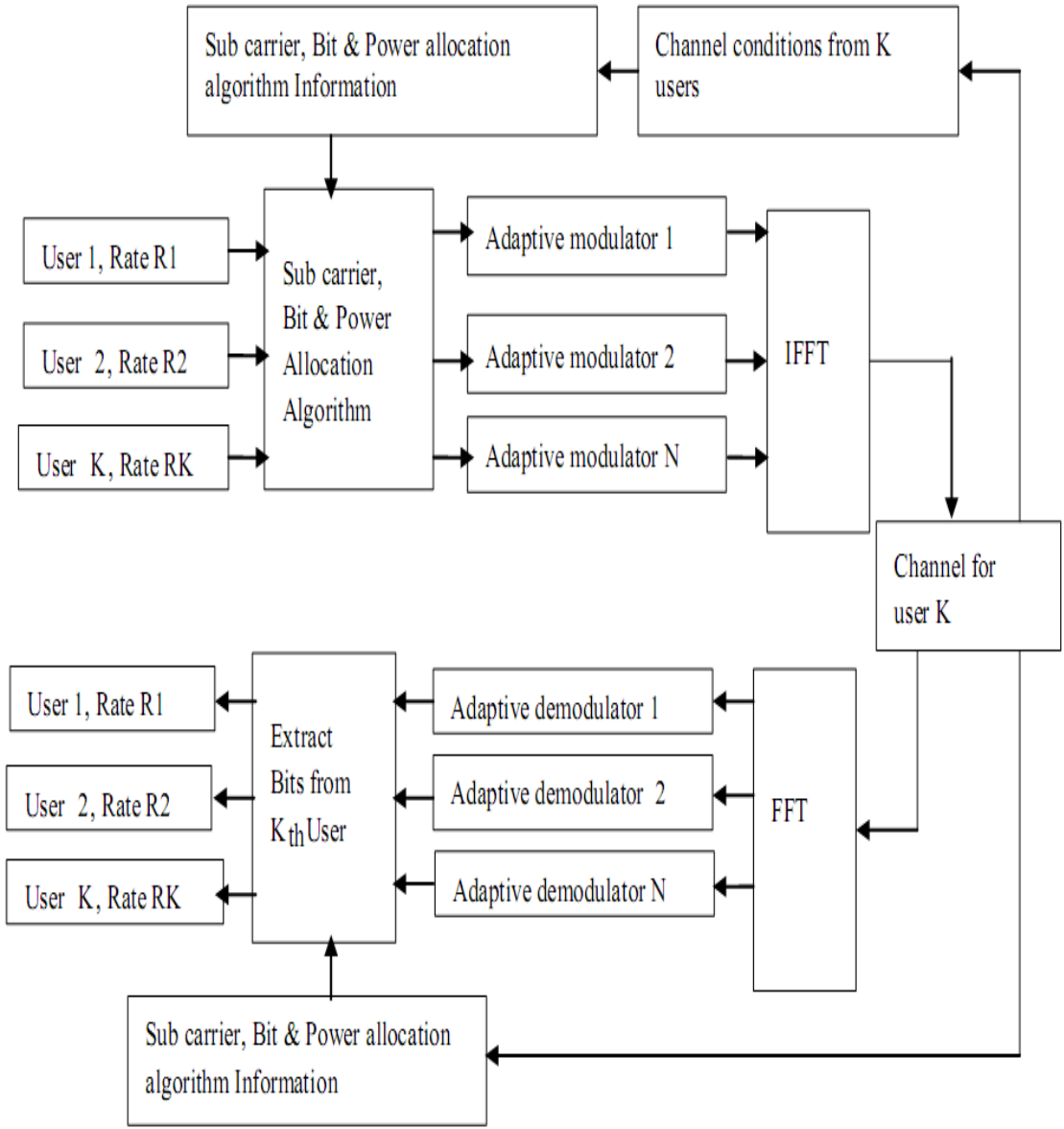

Figure 1: System model for multiuser OFDMA system [6]

This information is used to configure the transmitter. At the receiver, the subcarrier and power allocation information is used to configure the subcarrier selector and decoder to extract the data from the subcarriers assigned to the users. Throughout this paper, certain variables are used to denote the parameters of resource allocation as explained below. It is assumed that a total of *K* users in the system sharing *N* sub channels with total transmit power constraint $P_{TOT}$. The main objective is to optimize the sub channel and power allocation to achieve the highest sum error-free capacity under the total power constraint. It is assumed that each user experiences independent fading and the channel gain of user k in subcarrier *n* is denoted as $h_{k,n}$, with additive white Gaussian noise (AWGN) $\sigma^2 = N_o \frac{B}{N}$ where $N_o$ is the noise power spectral density. The corresponding sub-channel-to-noise ratio is thus denoted as $H_{k,n} = h_{k,n}^2/\sigma^2$.

# 4. RESOURCE ALLOCATION ALGORITHMS FOR OFDMA SYSTEMS

To maximize the total capacity, the transmission power needs to be distributed optimally under the total power constraint. The following optimization problem is formulated to determine the optimal power allocation. Mathematically, the optimization problem considered in this paper is formulated as[7]

$$\max \rho_{k,n} \sum_{k=1}^{k} \sum_{n=1}^{N} \frac{\rho_{k,n}}{N} \log_2 \left[ 1 + \frac{P_{k,n} h_{k,n}^2}{N_o \frac{B}{N}} \right] \quad (1)$$

Subject to following constraints

$$\sum_{k=1}^{k} \sum_{n=1}^{N} p_{k,n} \leq P_{TOT}$$

$$p_{k,n} \geq 0 \forall k, n$$

$$p_{k,n} = \{0,1\} \forall k, n$$

$$\sum_{k=1}^{k} p_{k,n} = 1 \forall n$$

where, $B$ is the total available bandwidth, $P_{k,n}$ is the power allocated for user k in the subchannel $n$, $\rho_{k,n}$ is either 1 or 0, indicating whether subchannel $n$ is used by user k or not. The capacity for user k, denoted as $R_k$, is defined as

$$R_k = \sum_{n=1}^{N} \frac{\rho_{k,n}}{N} \log_2 \left[ 1 + \frac{p_{k,n} h_{k,n}^2}{N_o \frac{B}{N}} \right] \quad (2)$$

## 4.1 Root-Finding Method of Subcarrier and Power Allocation

The root-finding method refers to the adaptive resource allocation proposed in. The name stems from a non-linear equation used to determine the total power assigned to the first user. In root-finding method, the determination of subchannel and power assignment was handled separately. This reduces the computational complexity of solving both variables simultaneously. The proportionality between the users is strictly maintained in this algorithm. A subchannel is initially assigned to the user who has the highest equalized channel gain in that subchannel. After all the subchannels are assigned, the capacity for each user is calculated, assuming equal power distribution. At each iteration, the user with the lowest proportional capacity has the option to pick which subchannel to use, provided the user should have not met their requirements. The subchannel allocation algorithm is suboptimal because equal power distribution in all subchannels is assumed. To assign power over all the subchannels, the root-finding method first determines the total power for each user and then breaks it down to the individual subchannels[8,9]. The total power of every user is calculated by using the equation

$$Pk, total = c_k(P_{1,tot})d_k \qquad (3)$$

The final step in the power allocation is to perform water filling across the subcarriers for each user

$$P_{K,N} = P_{K,1} + \frac{H_{k,n} - H_{k,1}}{H_{k,n} H_{k,1}} \qquad (4)$$

## 4.2 Linear Method of Subcarrier and Power Allocation

Wong *et al* algorithm is almost similar to Shen *et al* . algorithm with minor difference such as use of unallocated subcarriers and low complexity power allocation algorithm. There are four steps in subchannel allocation algorithm. Step 1 determines the number of subcarriers $N_k$ to be initially assigned to each user. step 2 assigns the subcarriers to each user with highest channel-to-noise ratio. Step 3 picks the user with lowest rate to allocate subchannel at the same time the proportionality between the users are roughly maintained. The step 4 assigns the remaining $N *$ subcarriers to the best users for them, wherein each user can get at most one unassigned subcarrier. This is to prevent the user with the best gains to get the rest of the subcarriers. This policy balances achieving proportional fairness while increasing overall capacity. The linear

$$Pk = \begin{cases} \dfrac{(Ptot - \sum_{k=2}^{k} \frac{bk}{akk})}{(1 - \sum_{k=2}^{k} \frac{1}{akk})} & for\ k \to 1 \\ \dfrac{(bk - P1)}{akk} & for\ k \to 2,3,\ldots.k \end{cases} \qquad (5)$$

equations involved in the power allocation can be easily solved because of its well-ordered symmetric and sparse structure. To solve the above equation, LU factorization (factorization of a matrix in to lower and upper triangular matrices of the same size respectively) is performed on the coefficient matrix. Finally, using forwards-backwards substitution, the individual user powers are given by The power of individual subchannel is found using the water-filling algorithm[10].

## 4.3 Joint Subcarrier and Power Allocation Algorithm

The algorithm simplifies the optimization problem into one that has *N* optimization parameters by assuming equal power allocation to all subcarriers, i.e

$$P_{k,n} = \begin{cases} \dfrac{Ptot}{N} & for \to n \in A \quad for\ all\ n = 1,2,\ldots,N \\ 0 \to otherwise & and\ k = 1,2,\ldots K \end{cases} \qquad (6)$$

In the joint allocation, optimization of the *N + K* parameters is carried out by alternating between subcarrier and power allocation. The water filling is used for each user. However, unlike, water filling for each user plays a crucial role in deciding the subcarrier allocation. When a subcarrier is allocated to a user, the power allocated to the user is incremented by P total/ *N*, i.e., the power allocated to each user is proportional to the number of subcarriers currently allocated to the user. The user's rate is also updated assuming that water filling is used. This updated rate information is used in the allocation of the remaining subcarriers[11,12].

# 5. BIT ALLOCATION ALGORITHM FOR OFDMA SYSTEMS

The algorithm in is aimed at the maximization of data rate under the constraint of total transmit power and BER. The algorithm is based on the water-filling approach, which is known as optimal to maximize data rate under the constraint of total transmit power. In the given algorithm, however, the water-filling power allocation is not fully performed. Instead, by adjusting only the water-filling level needed in the water-filling power allocation, we may adapt transmit power and number of bits for each subchannel with low computational complexity, and the total number of loaded bits in an OFDM symbol can be maximized while satisfying the constraints. The number of loaded bits, $b_m$ for the $m^{th}$ subchannel may be obtained by

$$b_m = \log_2\left(1 + \frac{S_m g_m}{\sigma^2 \Gamma}\right) \qquad (7)$$

$$\tilde{b}_m = round(b_m) \qquad (8)$$

where, $S_m$ and $g_m$ denote the transmit power allocated to the $m^{th}$ subchannel, and the power gain of the $m^{th}$ subchannel, respectively. $\sigma$ is the variance of the additive white Gaussian noise. $\Gamma$ represents the SNR gap which is a function of the target BER and a channel coding scheme. Note that the number of bits for each subchannel increases with $\lambda$, as seen from

$$b_m = [\log_2(\lambda g_m)]^+ \qquad (9)$$

where $[x]^+ = \max\{x, 0\}$. The increase of $b_m$ with $\lambda$ indicates the increase of $\tilde{b}$ and $\hat{s}$ as well from (7). Therefore, to maximize the data rate of the system, the water-filling level $\lambda$ should be kept as large as possible while satisfying the total transmit power constraint. Consequently, by adjusting $\lambda$, transmit power and number of bits for each subchannel can be adapted iteratively, and the number of loaded bits in an OFDM symbol can be maximized while satisfying the total transmit power constraint and BER requirement. At each iteration, the water-filling level $\lambda$ is adjusted as

$$\lambda = \lambda + \mu \frac{1}{M_{ON}} \cdot \frac{1}{\sigma^2 \Gamma}\left(S - \sum_{m=1}^{M} s_m\right) \qquad (10)$$

where $\mu$ is a step size; $M_{on}$ denotes the number of turned-on subchannels which are allowed to transmit data bits. The transmit power for the $m^{th}$ subchannel is calculated as

$$S_m = \frac{\sigma^2 \Gamma}{g_m}(2^{b_m} - 1) \qquad (11)$$

$b_m$ and $\tilde{b}_m$ are determined by only the value of the water-filling level, $\lambda$, when the channel power gain is available. Therefore, whole water-filling power allocation is not required as in, but only the value of $\lambda$ is adjusted to satisfy the total transmit power constraint given as

$$\sum_{m=1}^{M} s_m \leq S.$$

# 6. PROPOSED RESOURCE ALLOCATION ALGORITHM FOR OFDMA SYSTEMS

## 6.1 Number of sub-carriers determination

- **Inputs**: Each user's bit rate constraint and average channel gain for each user
- **Output**: Number of sub-carriers each user gets assigned
- Two types of sub-carriers: Minimum required sub-carrier and Extra sub-carrier
- Minimum required sub-carriers are to fulfil the user's bit rate constraint in the case that maximum amount of bits will be transmitted in each sub-carrier
- Extra sub-carriers will share bits with minimum required sub-carriers so that the loaded bits in each sub-carrier can be reduced and with an adaptive modulation scheme transmission power to all user can decrease
- No free sub-carrier left

## 6.2 Sub-carrier allocation

- Inputs: Channel State Information for each user and number of sub-carriers each user gets assigned
- Output: sub-carriers allocation

Phase 1: Constructive initial allocation

1. List the sub-carriers for each user in descend order according to channel gain
2. Check sub-carriers user by user if the number of sub-carrier each user gets is achieved or the sub-carrier has already been assigned to some users
3. If both are NO, assign the sub-carrier to this user, otherwise skip this user to next user

Phase 1 may achieve only a local minimum but not total minimum transmission power

Phase 2: Iterative improvement
- For every iteration, swap a pair of sub-carriers allocated to two users such that the result power can be reduced further
- Power reduction factor is the cost function in order to select the pair of users and pair of sub-carriers which can reduce power most
- Iteration is over when the maximum possible power reduction is less than zero

## 6.3 Bit loading

- Inputs: Sub-carriers allocation, channel gain and bit rate constraint
- Output: Bits loaded to achieve each user's bit rate constraint
1. Each time selecting the sub-carrier that requires the least additional power to add one more bit
2. Check if the maximum amount of bits loaded in this sub-carrier has already been achieved and if this user's bit rate constraint has been fulfilled
3. If both are NO, loading one more bit to this sub-carrier, otherwise selecting the sub-carrier which requires second least additional power and repeat 2

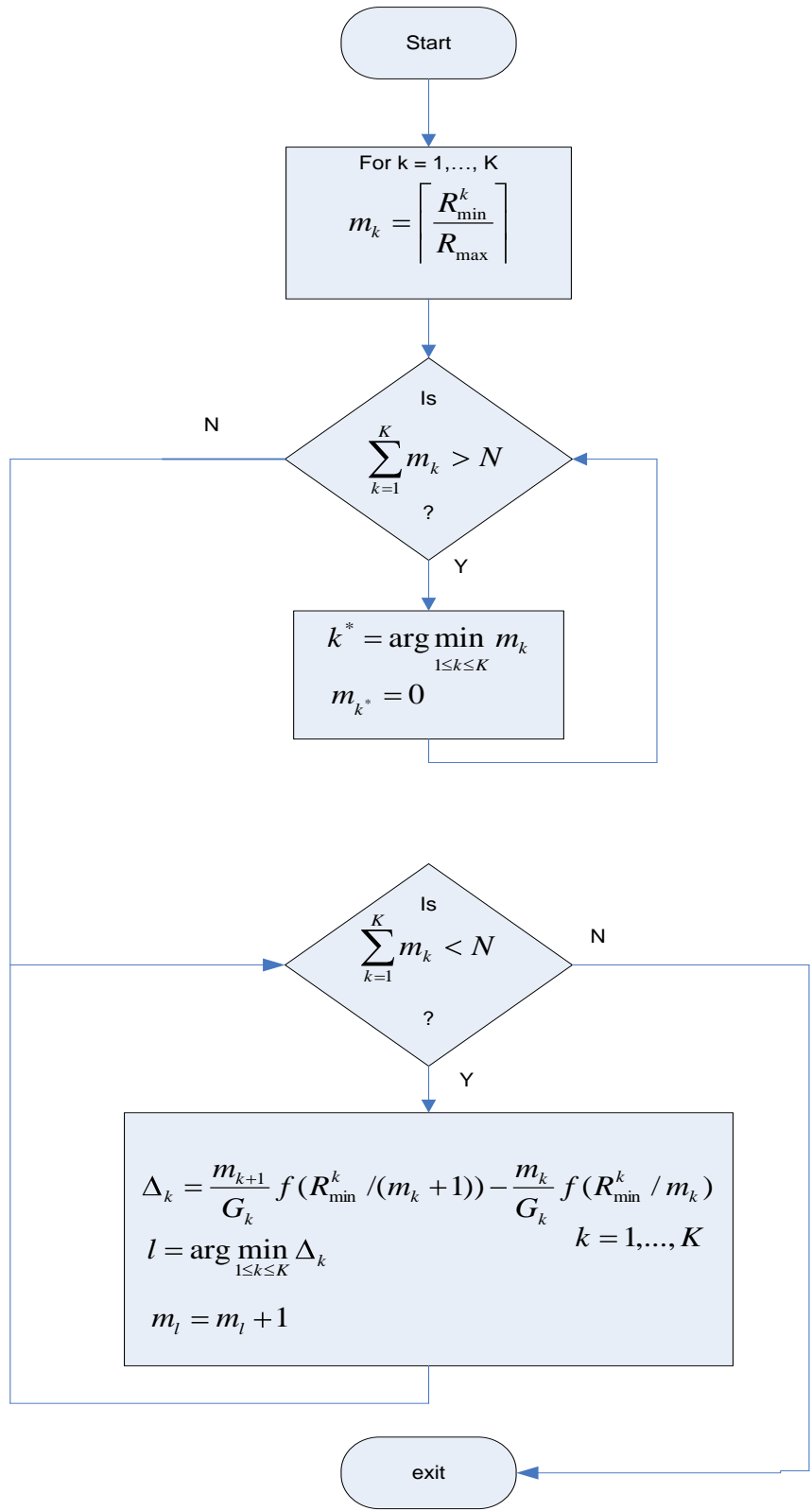

Figure 2 Subcarrier determination

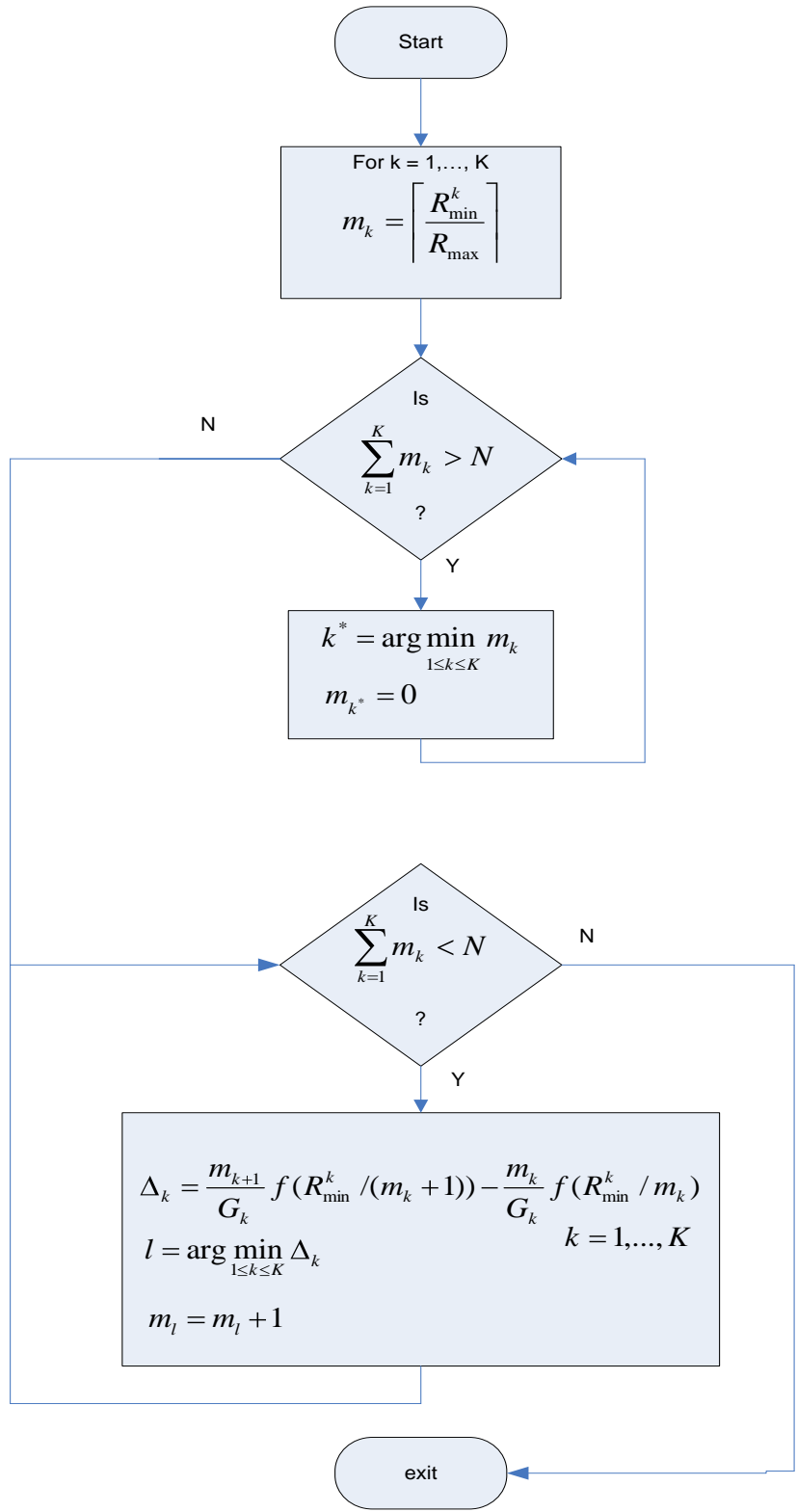

Figure 3 Subcarrier allocation (constructive initial allocation)

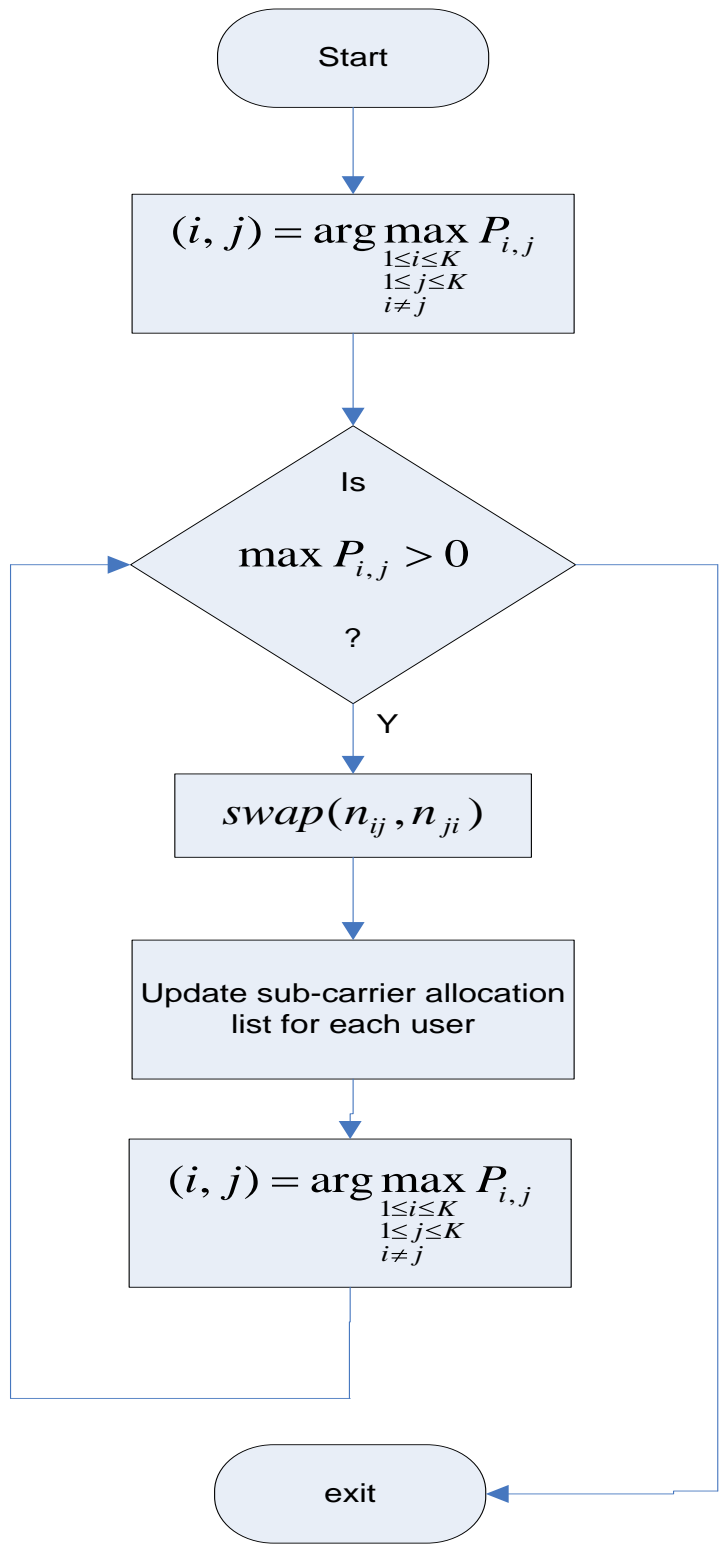

Figure 4 Subcarrier allocation(Iterative improvement)

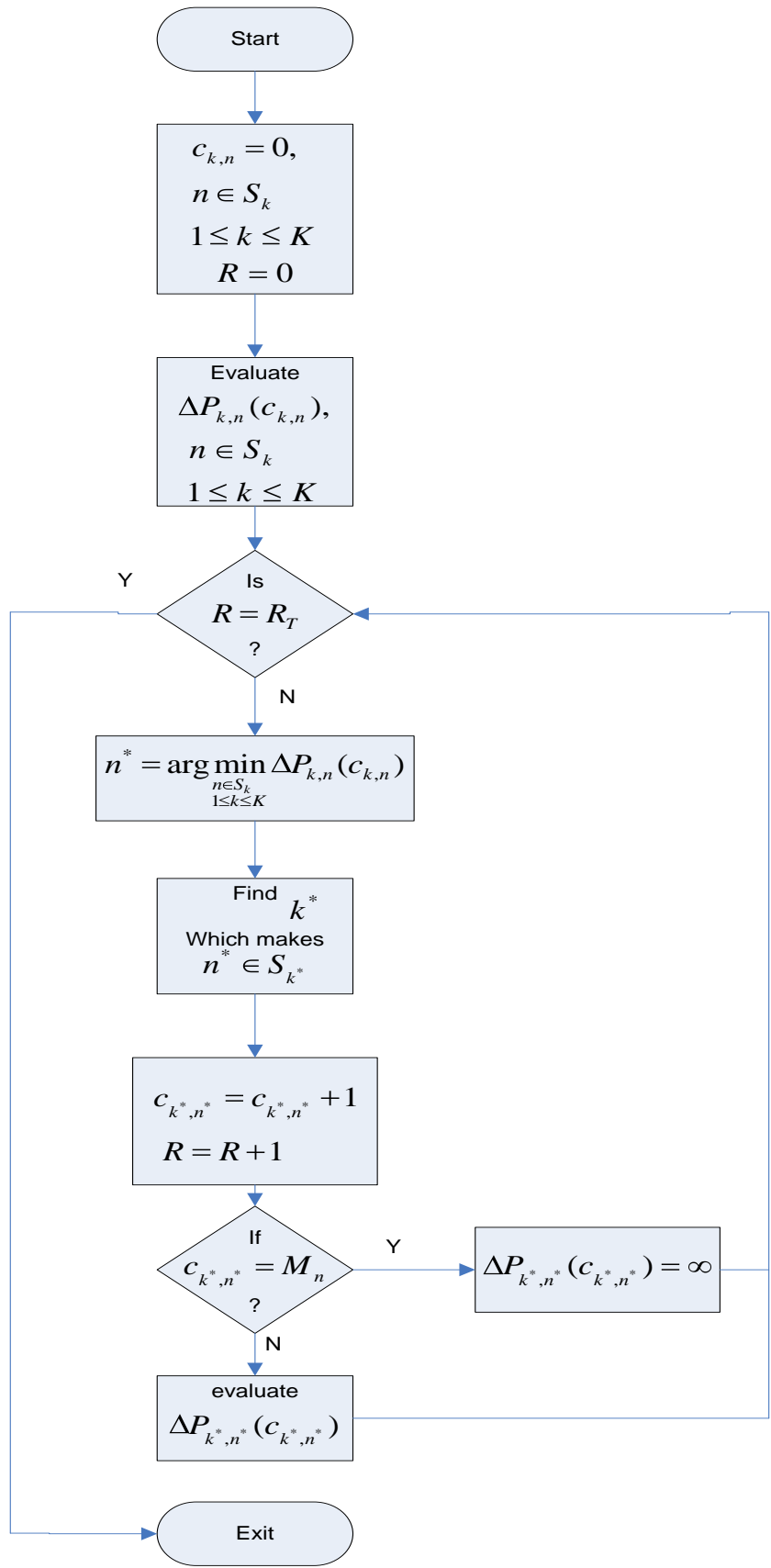

Figure 5 Bit loading

### 6.4 Primitive parameters

The following four primitive parameters characterize the OFDMA symbol:
*BW: This* is the nominal channel bandwidth.
*N*used: Number of used subcarriers (which includes the DC subcarrier).
*n*:Sampling factor. This parameter, in conjunction with *BW* and *N*used determines the subcarrier spacing, and the useful symbol time. This value is set to 8/7.
*G*:This is the ratio of CP time to "useful" time. The following values shall be supported: 1/32, 1/16, 1/8, and 1/4.

### 6.5 Derived parameters

The following parameters are defined in terms of the primitive parameters of
Smallest power of two greater than Nused
Sampling Frequency: $Fs = floor(n \cdot BW/8000) \times 8000$
Subcarrier spacing: $\Delta f = Fs/NFFT$
Useful symbol time: $Tb = 1/\Delta f$
CP Time: $Tg = G \cdot Tb$
OFDMA Symbol Time: $Ts = Tb + Tg$
Sampling time: $Tb/NFFT$

## 7. RESULTS AND DISCUSSION

Figure 6 shows the results which are calculated for 64 subcarriers and subchannel SNR = 38 dB and SNR gap ¡ = 3:3.The capacity achieved by the proposed LINEAR method is consistently higher than for the NONLINEAR method for a system with 4 up to 16 users. Figure 6 shows the comparison of total capacities between the proposed LINEAR method and NONLINEAR method . Notice that the capacities increase as the number of users increases. This is the effect of multiuser diversity gain, which is more prominent in systems with larger number of users. The proposed LINEAR method has a consistently higher total capacity than the NON- LINEAR method for all the numbers of users for this set. This analysis is done in Matlab R © 2008a

Figure 7 gives the Normalized capacity ratios per user for 16 users averaged over 100 channels, with the required Gamma shown as the leftmost bar for each user. The LINEAR method has minimal deviation from the required proportions, but Proposed scheme is much better for large number of users .The normalized proportions of the capacities for each user for the case of 16 users averaged over 100 channel samples

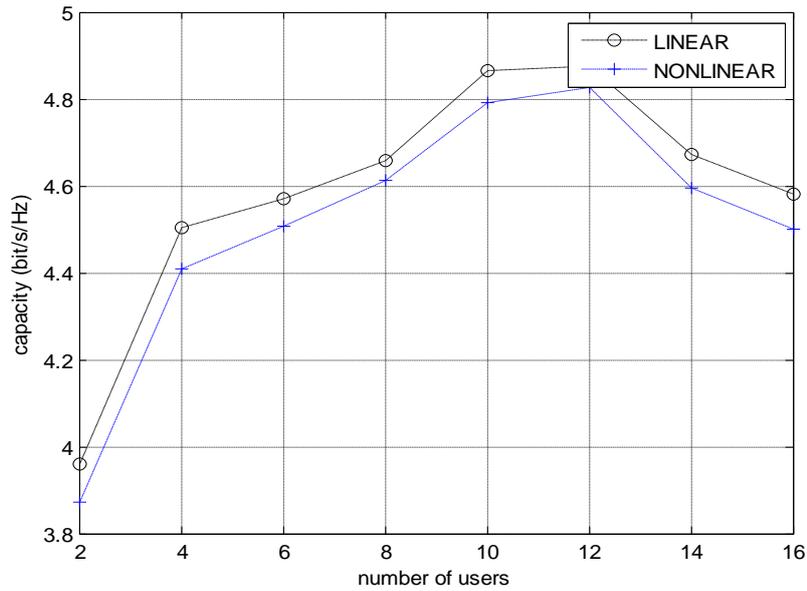

Figure 6 Total capacity versus number of users in an OFDMA system

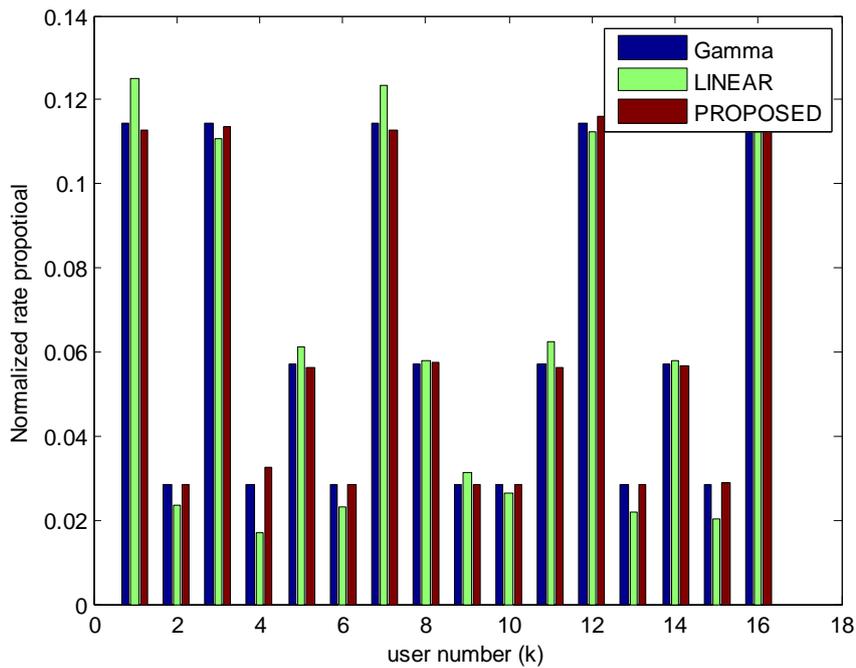

Figure 7 Normalized capacity ratios per user

## 8. CONCLSION

This paper presents a new method to solve the rate-adaptive resource allocation problem with proportional rate constraints for OFDMA systems. It improves on the previous work in this area by developing a novel subcarrier allocation scheme that achieves approximate rate proportionality while maximizing the total capacity. It is shown through simulation that the proposed method performs better than the previous work in terms of significantly decreasing the computational complexity, and yet achieving higher total capacities, while being applicable to a more general class of systems.

## REFERENCES

[1]     P. Parag, S. Bhashyam, and R. Aravind, "A subcarrier allocation algorithm for OFDMA using buffer and channel state information," in Proc. 62nd IEEE Veh. Technol. Conf., Sep. 2005, vol. 1, pp. 622–625.
[2]     G. Song, Y. Li, L. J. Cimini, and H. Zheng, "Joint channel-aware and queue-aware data scheduling in multiple shared wireless channels," in Proc. IEEE WCNC, Mar. 2004, vol. 3, pp. 1939–1944
[3]     Zukang Shen, J. G. Andrews, and B. L. Evans, "Adaptive resource allocation in multiuser OFDM systems with proportional rate constraints," *IEEE Transactions on Wireless Communications*, Vol. 4, pp. 2726 2737, 2005.
[4]     I. C. Wong, Z. Shen, J. G. Andrews, and B. L. Evans, "A low complexity algorithm for proportional resource allocation in OFDMA systems," *Proc. IEEE Int. Work. signal Processing Systems*, pp. 1-6, 2004.
[5]     Didem Kivanc, Guoqin g Li, and H ui Liu, "Comput ationally Efficient B and width Allocation and Power Control for O FDMA," *IEEE Trans. Wireless Communication.*, vol. 2, pp. 1150-1158, Nov. 2003.
[6]     Must afa Er gen, S inem C oleri, and Pravin Varaiya , "QoS Awar e Adapti ve R esource Allocation Techniques for Fair Scheduling in OFDMA Based Broadband Wireless Access S ystems," *IEEE Trans. Broadcasting*, vol. 49, pp. 362-370, Dec. 2003.
[7]     Tolga Girici, Chenxi Zhu, Jonathan R. Agre, and Anthony Ephremides,"Proportional Fair Scheduling Algorithm in OFDMA Based Wireless Systems with QoS Constraints" journal of communications and networks, vol. 12, no. 1, February 2010
[8]     Alessandro Biagioni, Romano Fantacci, Dania Marabissi and Daniele Tarchi, "Adaptive Subcarrier Allocation Schemes for Wireless OFDMA Systems in WiMAX Networks". IEEE Journal on selected areas in communications, vol.27, No.2, FEB – 2009.
[9]     Kibeom S eong, M ehdi M ohseni, and J ohn M. C ioffi, "Optim al R esource Allocati on for OFDMA Downli nk S ystems," *Proc. IEEE Int'l S ymp. on Info. T heor y (ISIT )* , pp. 1394-1398, July 2006.
[10]    Mohmmad Anas, Kanghee Kim, Seokjoo Shin, and Kiseon Kim, "QoS aware power allocation for combined guaranteed performance and best effort users in OFDMA systems," *IEEE International Symposium on Intelligent Signal Processing and Communication Systems* ,pp. 477-481, 2004.
[11].    Jiho Jang, Kwang Bok Lee, and Yong-Hwan Lee, "Transmit power and bit allocations for OFDM systems in a fading channel," in *Proc. IEEE Global Communications Conference,* Vol. 2, pp. 858-862, 2003.
[12]    H. Yin and H. Liu, An Ef_cient Multiuser Loading Algorithm for OFDM-based Broadband Wireless Systems,î in *Proc. IEEE Global Telecommunications Conference*, vol. 1, 2000, pp. 103 to 107.
[13]    W. Rhee and J. M. Ciof_, Increase in Capacity of Multiuser OFDM System Using Dynamic Subchannel Allocation, in *Proc. IEEE Vehic. Tech. Conf.*, Tokyo, Japan, May 2000, pp. 1085 to 1089.
[14]    Z. Shen, J. G. Andrews, and B. L. Evans, Optimal Power Allocation in Multiuser OFDM Systems, in *Proc. IEEE Global Communications Conference*, San Francisco, CA, Dec. 2003, pp. 337 to 341.
[15]    S. T. Chung and A. Goldsmith, ìDegrees of Freedom in Adaptive Modulation: A Unied View, *IEEE Trans. Commun.*, vol. 49, no. 9, pp. 1561 to 1571, Sept 2001.
[16]    Z. Shen, J. G. Andrews, and B. L. Evans, Adaptive Resource Allocation in Multiuser OFDM Systems with Proportional Fairness, *IEEE Trans. Wireless Commun.*, Nov.2005 issue 6 pp 2726-2737.
[17]    Guobin Zhang, Suili Feng, "Subcarrier Allocation Algorithms Based on Graph Coloring in Cognitive Radio NC-OFDM System" South China University of Technology, Guangzhou, China Computer Science and Information Technology (ICCSIT), 2010 3[rd] IEEE International conference pp 535 to 540
[18]    IEEE Standard for Local and metropolitan area networks Part 16: Air Interface for Fixed Broadband Wireless Access Systems 802.16, 2004 pp 1-893.

**Authors**

**First Author** He obtained his BE degree in Electronics Engg from Kavikulguru Institute of Technology and Science, Ramtek, Nagpur University in 1996. M. Tech. degree in Telecommunication Engineering from Indian Institute of Technology, Kharagpur in 2002. Pursuing Ph.D. from R.T.M. Nagpur University Nagpur He has more than 15 years of teaching experience. He has published more than 25 research papers which include international journals, international conference, national journals & national conference. Presently he is working as a Asst. Prof in Electronics & Communication Dept at KITS Ramtek. His research interest includes multimedia over wireless, 4G mobile communication & Channel model for LTE, Resource allocation in OFDMA. He is member of IETE & ISTE & Life member of IE. Awarded with who's who in the world by marquis for the year Nov. 2009.

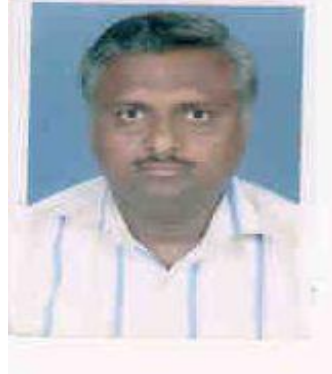

**Second Author** He born in the year 1960 and presently working as a principal ITM College of engineering, Kamptee Nagpur. He was elected as a Hon. Secretary of Institution of engineers, Nagpur local centre in 2008. He worked earlier from 2006 to 2008 as a professor, head of electronics and power engineering department and M.Tech Co-ordinator at Yeshwantrao Chavhan college of engineering, Nagpur. He worked as a principal twice during 1992 to 1994 and 1999 to 2001 at B.D.C.O. Sewagram. He was fully geared to implement the teaching programme effectively and efficiently to ensure quality education in institution. He was instrumental for academic and administrative growth of the B.D.C.O.E. Sewagram. He also elected as a chairman, board of studies in electronics engineering & member of Academic council of Nagpur University during 1995-2000 and 2003 to 2005.

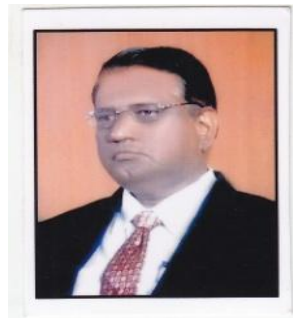

**Third Author** Dr. Rajesh Pande obtained his BE degree in Electronics Engineering from Govt. College of Engineering . Jabalpur in 1990. M. Tech. degree in Electronics Engineering from VRCE Nagpur in 1992. Ph.D. from V.N.I.T. Nagpur in 2008. He has more than 20 years of teaching experience. He has published more than 25 research papers which include international journals, international conference, & national conference, Presently he is working as a Prof. Electronics & Vice-Principal at Ramdeobaba College of Engineering & Management Nagpur. He is member of IEEE , ISTE & AMIE. His area of interest includes VLSI design, RFMEMS & Communication sysytems

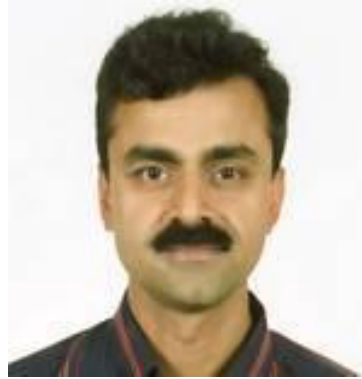

**Fourth Author** S.S.Pathak received his B.Tech and M.Tech degrees in Electronics Engineering from I.T.B.H.U. is 1976 and 1978 respectively. He got his Ph.D degree on Digital

Communications awarded in 1984 from I.I.T.Delhi. He has to his credit around 20 Journal publications and about 80 publications in conference proceedings. His area of research interest includes error control coding, Telecommunication switching networks, Wireless networks, Spectral efficiency of transmitted signals, and computer communication networks. He joined I.I.T. Kharagpur as a lecturer in 1985 where he is currently associated with the Department of Electronics and Electrical Communication Engineering as Professor.

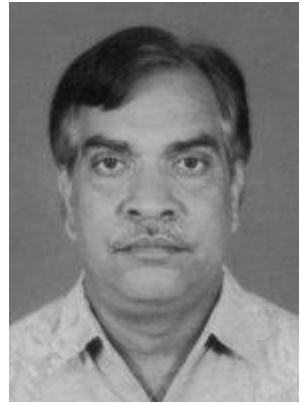

.